\documentclass[conference]{IEEEtran}
\IEEEoverridecommandlockouts
\usepackage{graphicx} 
\usepackage{pifont}
\usepackage{amsmath}
\usepackage{amssymb}  
\usepackage{amsfonts}
\usepackage{algorithmic}
\usepackage{array}
\usepackage{multicol}
\usepackage{svg}
\usepackage{textcomp}
\usepackage{stfloats}
\usepackage{url}
\usepackage{verbatim}
\usepackage{graphicx}
\usepackage{caption}
\usepackage{subcaption}
\usepackage[ruled, vlined, linesnumbered]{algorithm2e}
\usepackage{cite}
\usepackage{booktabs}
\usepackage{amssymb,amsfonts,bm}
\usepackage{amsmath,tikz}
\usepackage{color}
\usepackage{mathtools}
\usepackage[hidelinks]{hyperref} 
\usepackage{rotating}
\usepackage{blkarray}
\usepackage{physics}
\usetikzlibrary{arrows}
\usepackage{makecell}

\usepackage{amsthm}
\usepackage{comment}
\usepackage{multirow}
\usepackage{geometry}
\graphicspath{{images/}}

\begin{document}

\title{\LARGE \bf

AV-DTEC: Self-Supervised Audio-Visual Fusion for Drone Trajectory Estimation and Classification\\

\author{Zhenyuan Xiao, Yizhuo Yang, Guili Xu*, Xianglong Zeng, Shenghai Yuan


\thanks{This research is supported by the Postgraduate Research and Practice Innovation Program of Jiangsu Province.}
\thanks{Zhenyuan Xiao, Guili Xu, and Xianglong Zeng are with the College of Automation Engineering, Nanhang University, China, 210000, 
   { Email: zy.xiao@nuaa.edu.cn, guilixu2002@163.com, zxl2024@nuaa.edu.cn}. Yizhuo Yang and Shenghai Yuan are with the School of Electrical
and Electronic Engineering, Nanyang Technological University, Singapore 639798, {Email: yang0670@e.ntu.edu.sg, shyuan@ntu.edu.sg} } 
}
}

\maketitle

\begin{abstract}
The increasing use of compact UAVs has created significant threats to public safety, while traditional drone detection systems are often bulky and costly. To address these challenges, we propose AV-DTEC, a lightweight self-supervised audio-visual fusion-based anti-UAV system. AV-DTEC is trained using self-supervised learning with labels generated by LiDAR, and it simultaneously learns audio and visual features through a parallel selective state-space model. With the learned features, a specially designed plug-and-play primary-auxiliary feature enhancement module integrates visual features into audio features for better robustness in cross-lighting conditions. To reduce reliance on auxiliary features and align modalities, we propose a teacher-student model that adaptively adjusts the weighting of visual features. AV-DTEC demonstrates exceptional accuracy and effectiveness in real-world multi-modality data. The code and trained models are publicly accessible on GitHub
 \url{https://github.com/AmazingDay1/AV-DETC}.

\end{abstract}

\section{Introduction}
In the last decades, unmanned aerial vehicles (UAVs) have made significant contributions to goods delivery, inspection, and entertainment with their ease of use and affordability ~\cite{gonzalez2017unmanned, mishra2020drone}.  However, these UAVs can become hidden dangers and be used for malicious intent~\cite{jevvcak2019mobile}, such as smuggling, air traffic distribution, etc.   
Effective solutions are needed to detect and mitigate these threats, as shown in Fig. \ref{fig0}.


Most existing methods rely on single-modal detectors, such as visual~\cite{zheng2021air_rgb_dtc}, audio~\cite{vora2023dronechase}, or LiDAR~\cite{deng2024multi_point}. However, single-modal methods have inherent limitations: visual failure under lighting change, audio can be disrupted by noise, and LiDAR struggles with obstruction. To address these issues, some approaches fuse multiple modalities, such as combining visual and audio~\cite{yang2023av_audio_img_fusion}, to compensate for these limitations. However, these methods require extensive annotation, which is challenging in real-world applications. In short, the challenge is finding an efficient way to fuse different modalities without requiring manual annotations.

In response to these challenges, we propose AV-DTEC, a self-supervised audio-visual fusion method for drone trajectory estimation and classification. AV-DTEC is mainly composed of a selective state-space model (SSM) ~\cite{gu2023mamba}. By leveraging the correspondence between modalities, our model achieves self-supervised training without annotation. 
At any given moment, different modalities can capture UAV signals simultaneously during training. We utilize the spatiotemporal relationships of different modalities within the model. During the testing phases, only a lower-cost modality is needed for deployment. 
Our contributions can be summarized as follows:

\begin{figure}[t]
\centering
\includegraphics[width=8.6cm]{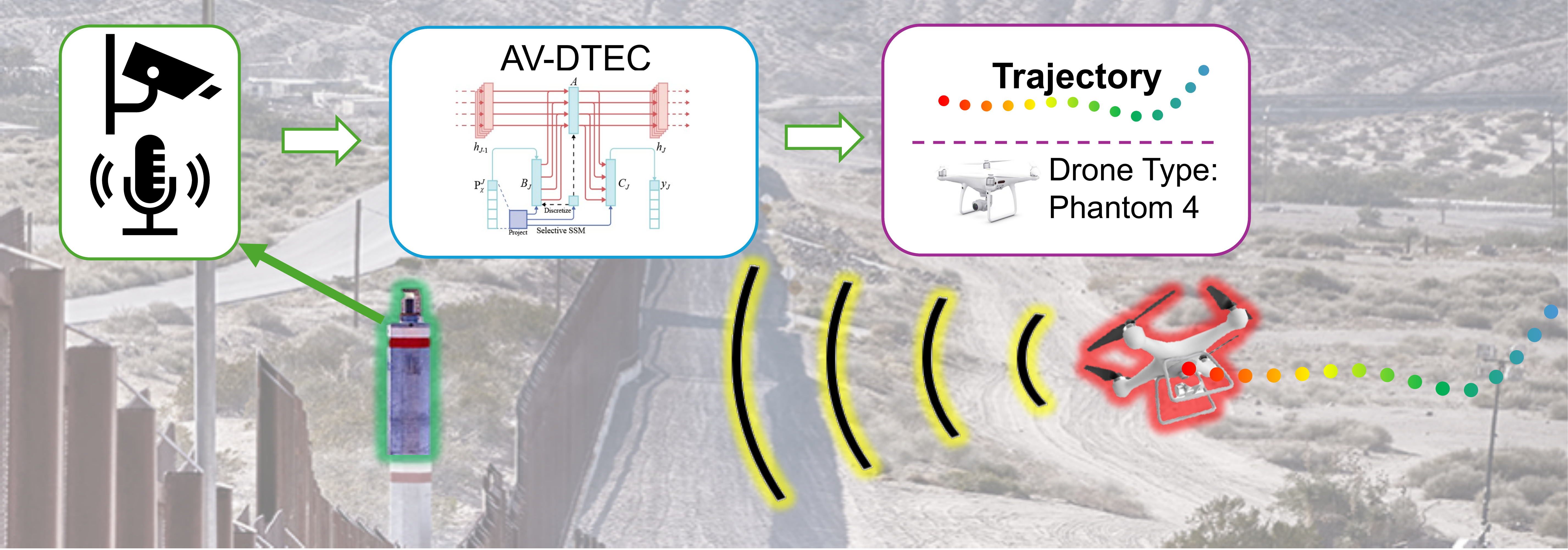}
\caption{Our audio-visual fusion model effectively identifies and locates drug-smuggling UAVs with high robustness and cost efficiency.}
\label{fig0} 
\vspace{-2em}
\end{figure}


\begin{itemize}
  \item We propose a novel anti-UAV system utilizing multi-modal fusion within a self-supervised framework. This approach eliminates annotation, notably combining visual and audio for hemispherical UAV trajectory estimation and classification.
  \item We introduce the Audio Mamba for audio feature extraction, employing learnable token learning of the temporal difference
of sound propagation.
  \item 
  We developed a plug-and-play feature enhancement module with an adaptive adjustment mechanism, integrating auxiliary features into primary ones via residual cross-attention, controlling the degree of fusion.
  \item We open-source our work for the benefit of the community. 
\end{itemize}


\vspace{-0.25em}
\section{Related Work}

\subsection{Anti-UAV Detection System}

In recent years, numerous anti-UAV systems and challenges have emerged to address the growing drone problem. The existing research primarily focuses on classification, 2D-tracking, and 2D-trajectory estimation. For instance, some institutions have compiled extensive RGB datasets, approaching the anti-UAV as an object detection~\cite{zhao2022vision_rgb_dtc_tracking,munir2024investigation_rgb_dtc,zheng2021air_rgb_dtc}. The three years of anti-UAV Challenge~\cite{jiang2021anti_1nd_IF,zhao21082nd_IF,zhao20233rd_IF} attracted scholars to explore 2D-object detection and 2D-tracking with infrared images. Several promising anti-UAV methods emerged from this competition, such as TAD~\cite{lyu2023real_IF_dtc}, which uses inconsistent motion between drones and backgrounds to identify potential drones. SiamDT tracker~\cite{huang2023anti_IF_tracking} employs a dual semantic feature extraction mechanism to enable effective target modeling and suppress background interference. Other researchers have explored the use of audio for drone classification~\cite{wang2022large_audio_cls,ali2024exploitinge_audio_cls}. However, these methods lack 3D trajectory. To address this, some scholars have proposed using point cloud features, trajectory smoothness, and motion characteristics for unsupervised UAV trajectory estimation~\cite{deng2024multi_point,wu2024vehicle_point,vrba2023onboard_point}. 
While these methods can output 3D trajectories, point clouds lack category information and are costly and difficult to set up. Additionally, the compact size of UAVs often results in sparse point clouds, making it challenging to differentiate between birds and UAVs.

\subsection{Audio-Visual Fusion Model}
Multi-modal fusion, especially of audio and visual data, mimics the human sensory system and can enhance intelligent system performance. This fusion, widely applied in object detection, improves spatial resolution and adds context to audio activity.
For instance, converting audio into a mel-spectrogram enhances its features, which are then fused with visual features to locate the sound source \cite{chen2020soundspaces_audio_navigation,tao2021someone,jiang2022egocentric_audio_img_location}. 
However, these methods heavily rely on visual assumptions and are ineffective in varying conditions, such as day and night.
DroneChase~\cite{vora2023dronechase} reduces visual reliance with a self-supervised framework, using pseudo labels from a visual backbone to train an audio-based network. However, it remains biased by line-of-sight conditions.
Subsequent self-supervised research on audio-visual fusion enabled object detection~\cite{yang2023av_audio_img_fusion} in both day and night conditions, effectively addressing moving objects \cite{valverde2021there_audio_car, gan2019self_audio_car}. Building on this, AV-FDTI~\cite{yang2024av_Audio-visual} adopted a supervised audio-visual fusion approach for anti-UAV, utilizing the cross-attention architecture of ViT \cite{dosovitskiy2020image_transformer} for fusion and 3D trajectory estimation and classification. However, AV-FDTI models are complex and require manual annotation. 


\vspace{-0.25em}
\section{Proposed Method}
\begin{figure*}[t]
\centering
\centering
\includegraphics[width=0.9\linewidth]{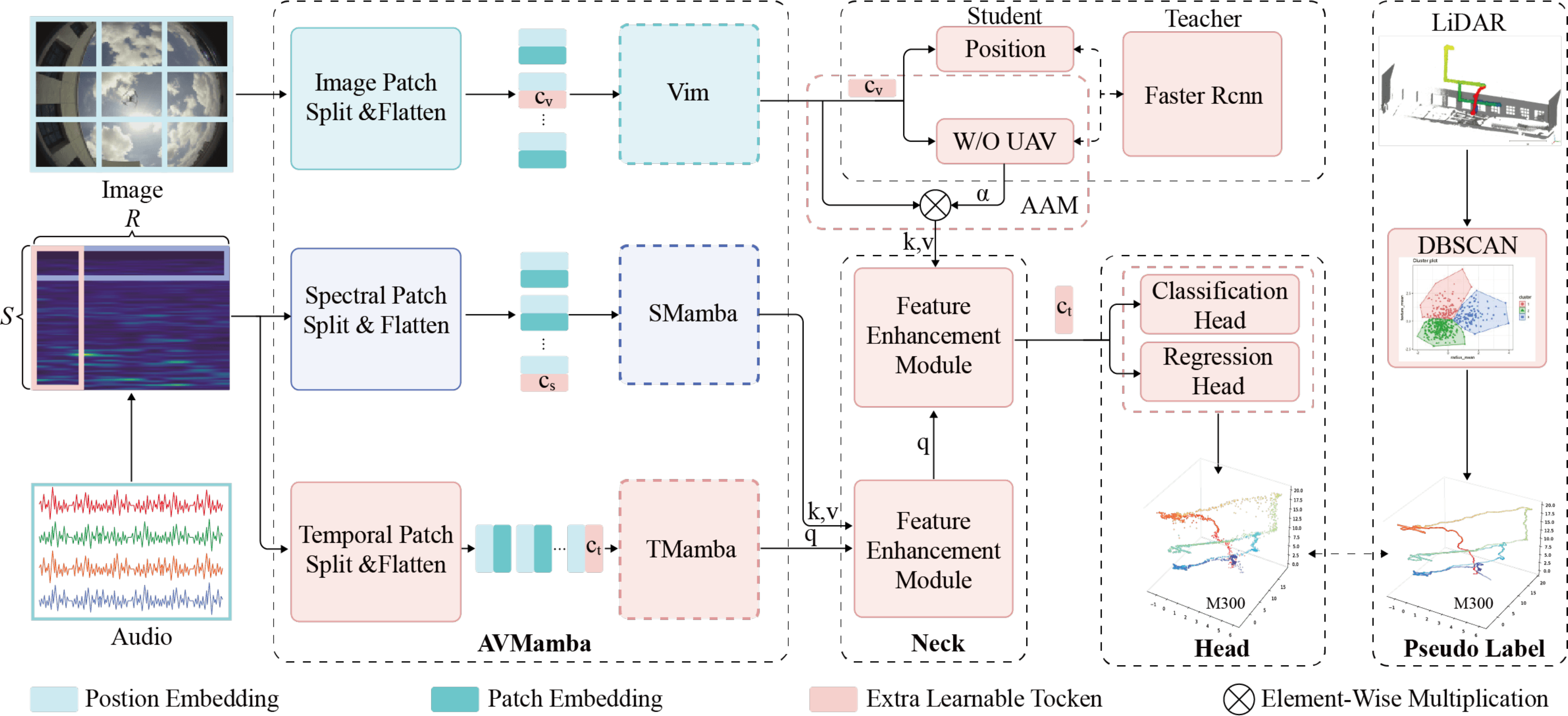}
\vspace{-0.5em}
\caption{AV-DTEC Architecture. During training, the learnable visual token extracted by Vim is trained through the teacher-student model to output the UAV center position and existence probability. For the inference, the token only outputs the existence probability, which is used to adjust the proportion of visual features.}
\label{structure} 
\vspace{-2em}
\end{figure*}

We propose a self-supervised audio-visual fusion system for UAV trajectory estimation and classification within a hemispherical area. In this system, AV-DTEC is trained with labels generated by LiDAR. As illustrated in Fig. \ref{structure}, the model comprises the Audio-Visual Mamba (AVMamba) feature extraction backbone, a feature fusion neck, and a prediction head. The AVMamba consists of Audio Mamba (AMamba) and Vision Mamba~\cite{zhu2024vision_mamba} (Vim) to extract audio and visual features. AMamba comprises Temporal Mamba (TMamba) and Spectral Mamba (SMamba), which extract the temporal difference of arrival and spectral attenuation. Vim is responsible for detecting the drone in images and mapping its 3D trajectory to 2D image positions. 
To fuse and align the auxiliary feature and primary feature, we designed a feature enhancement module and an adaptive adjustment mechanism (AAM) that generates adjustment factors through the teacher-teacher model. 
The details of each module in the network will be elaborated in the following sections.

\subsection{Audio-Visual Mamba} \label{sec3.B}

\begin{figure}[t]
\centering
\includegraphics[width=8.6cm]{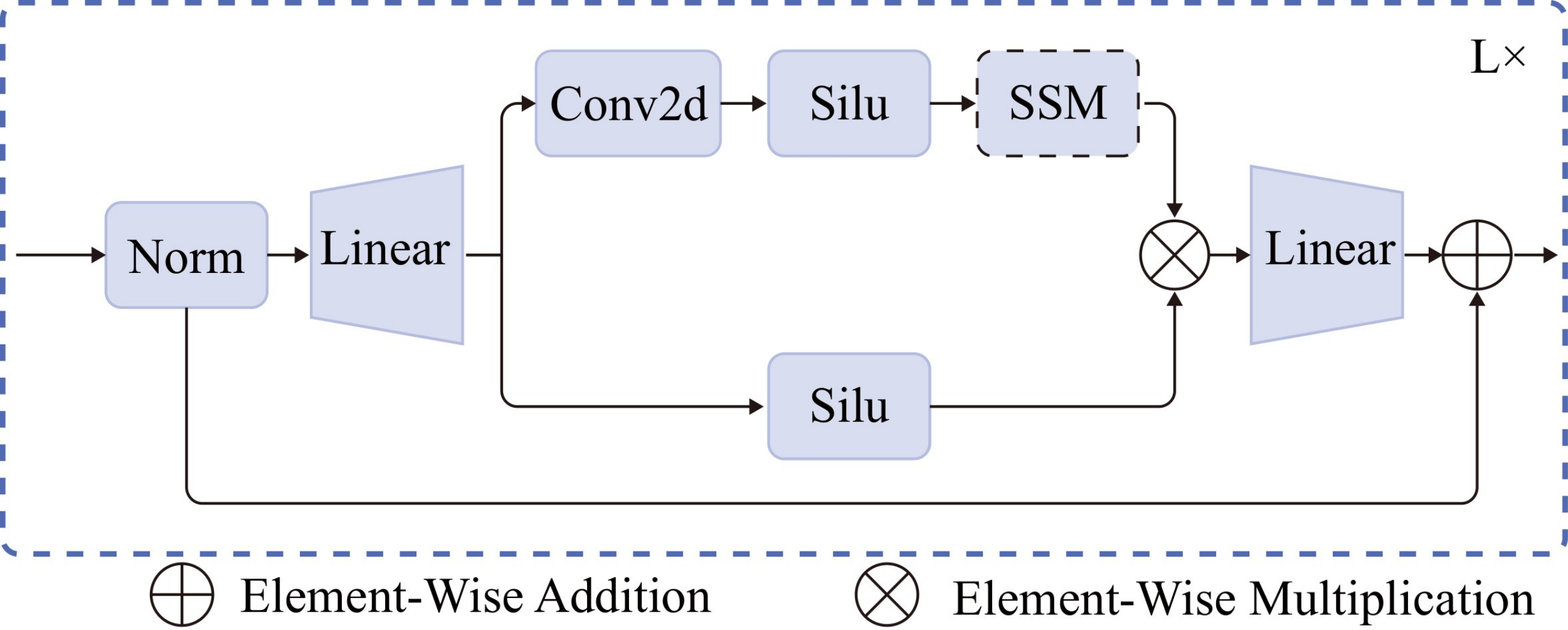}
\vspace{-1.5em}
\caption{The architecture of TMamba and SMamba block.}
\label{block} 
 \vspace{-2em}
\end{figure}

Audio features are extracted using the AMamba model, which consists of TMamba and SMamba. As illustrated in Fig. \ref{block}, TMamba and SMamba share an identical block. The input audio feature varies between TMamba and SMamba. The local and global differences in sound propagation are extracted through CNN and SSM. 

We define $\zeta$ as the number of microphone channels, \textit{R} as the width length of the spectrogram, \textit{S} as its height, and \textit{w} and \textit{h} as the width and height of splitting patch, respectively. To extract these features, multi-channel audio is transformed into mel-spectrograms, denoted as $\boldsymbol{\textbf{U}} \in \mathbb{R}^{\zeta\times R \times S}$, which serve as input to AV-DTEC. Let ${\chi}$ represent either the temporal or spectral dimension. 
The spectrogram is divided and flattened along both the temporal and spectral axes to form the patch sequences $\textbf{p}_{\chi}$. Specifically, the temporal patch sequence is $\textbf{p}_{R} \in \mathbb{R}^{J\times (\zeta wS)}$, and the spectral patch sequence is ${\textbf{p}}_{S} \in \mathbb{R}^{J\times (\zeta Rh)}$.
Inspired by ViT~\cite{dosovitskiy2020image_transformer}, learnable tokens ${\textbf{t}}_{\chi}$ are added in both the temporal and spectral dimensions. The patch sequences ${\textbf{p}}_{\chi}$ are then projected into vectors of dimension \textit{D}, and positional embeddings ${\textbf{E}}_{pos} \in \mathbb{R}^{(J+1)\times D}$ are incorporated. The final patch embedding $\textbf{P}_{\chi}$ is obtained through linear projection. As follows:

\vspace{-1em} 
\begin{equation} 
\textbf{P}_{\chi}=[\textbf{p}_{\chi}^{1}\textbf{W};\textbf{p}_{\chi}^{2}\textbf{W};...;\textbf{p}_{\chi}^{J}\textbf{W};\textbf{t}_{\chi}]+\textbf{E}_{pos},
\end{equation}
\vspace{-1.5em}
 
 Meanwhile, non-overlapping convolutions are utilized to split the spectrogram into patches. During the temporal patch split (horizontal), the height of each patch remains consistent with the height of the spectrogram. TMamba processes the patch by scanning from left to right along the temporal axis. It captures the temporal of sound propagation. For the spectral patch split (vertical), the width of each patch is preserved as the full width of the spectrogram. SMamba scans from top to bottom along the spectral axis. This vertical scanning captures global spectral attenuation, allowing the model to extract features that describe the overall spectral characteristics.

TMamba and SMamba block is composed of SSM. The process is described as follows:

\vspace{-0.5em} 
\begin{equation} 
\begin{matrix}
\begin{aligned}
\textbf{h}_{J}&=\overline{\textbf{A}}\textbf{h}_{J-1}+\overline{\textbf{B}}\textbf{P}_{\chi}^{J}
 \\ \textbf{y}_{J}&=\textbf{C}\textbf{h}_{J} \\

\overline{\textbf{A}}&=exp(\varDelta \textbf{A})
 \\ \overline{\textbf{B}}&={exp(\varDelta \textbf{A})}^{-1}(exp(\varDelta \textbf{A})-I)\varDelta \textbf{B},
\end{aligned}
\end{matrix}
\end{equation}
\vspace{-0.5em}

Where ${\textbf{h}\in \mathbb{R}^{\textit{D}\times \textit{d}}}$ represent a summary of spectral attenuation features or temporal difference of arrival features. $\textbf{P}_{\chi}^{J}$ is the input for the \textit{J}th patch embedding. $\overline{\textbf{A}}$ and $\overline{\textbf{B}} \in \mathbb{R}^{\textit{D}\times \textit{d}}$ are learnable parameters obtained by discretizing the continuous parameters \textbf{A} and \textbf{B} using the time step $\varDelta$.  $\varDelta$, ${\textbf{C}\in \mathbb{R}^{\textit{D}\times \textit{d}}}$ and $\overline{\textbf{B}}$ adapt to the input $\textbf{P}_{\chi}$, capturing variations in audio feature across different patches.
$\textbf{h}_{0}$ is initialized to 0, representing no historical audio at the beginning. $\overline{\textbf{B}}$ filters the input features of each sequence patch. $\overline{\textbf{A}}$  adjusts and fuses the historical features $\textbf{h}_{J}$. \textbf{C} enhances the accumulated temporal or spectral feature to produce the output $\textbf{y}_{J}$, summarizing all previous patch features.


For visual feature extraction, we employ Vim directly and add classification and regression tasks in Vim with extra learnable visual tokens. It can be used to adaptively scale visual features and align audio and visual features. The description will be introduced in Sec ~\ref{secC.2}.

\subsection{Feature Fusion Neck}   \label{sec3.C}
\subsubsection{\textbf{Feature Enhancement Module}}

Audio, unlike visual features, enables detection under any lighting condition, making it more reliable for both day and night. Thus, audio should serve as the primary feature when fusing. In the context of audio, temporal and spectral features have similar properties. Temporal features, which are tied to sound propagation and correlate with distance, should be prioritized since they encode both spatial location and category information. 
Spectral features, on the other hand, capture the overall spectral distribution over time, also encoding less spatial location and category information. Ablation studies in \ref{sec4.ablation} reveal that temporal and spectral features are correlated, and concatenating them directly introduces redundant information, which can degrade performance.

\begin{figure}[t]
\centering
\includegraphics[width=8cm]{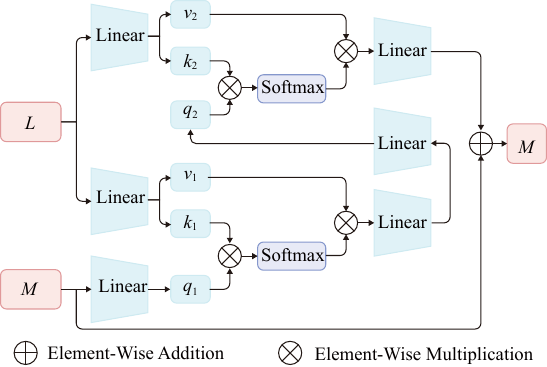}
\vspace{-0.5em}
\caption{Feature Enhancement Module.}
\label{fem} 
 \vspace{-2em}
\end{figure}

To address these, we designed a feature enhancement module (FEM), denoted as $\Phi$. As illustrated in Fig. \ref{fem}, the FEM leverages residual cross-attention, integrating auxiliary feature \textbf{L} into the primary feature \textbf{M}. The computation of FEM is performed as follows:

\vspace{-1.5em} 
\begin{equation} 
\begin{matrix}
\begin{aligned}

\text{$\Phi$}(\textbf{M},\textbf{L})=\textbf{M} +  \text{Attention}({{q}}_{2},\textbf{L}{\textbf{W}}^{{k}_{2}},\textbf{L}{\textbf{W}}^{{v}_{2}}){\textbf{W}}^{}

 \\ {q}_{2}= \text{Attention}(\textbf{M}{\textbf{W}}^{{q}_{1}},\textbf{L}{\textbf{W}}^{{k}_{1}},\textbf{L}{\textbf{W}}^{{v}_{1}})

 \\\text{Attention}(\textbf{Q},\textbf{K},\textbf{V})=\text{softmax}(\frac{\textbf{Q}{\textbf{K}}^{\text{T}}}{\sqrt{{d}_{k}}})\textbf{V},
\end{aligned}
\end{matrix}
\end{equation}
\vspace{-1.em} 

Where $\textbf{W}\in {\mathbb{R}}^{{d}_{m}\times {d}_{k}}$  represents learnable parameter matrices. $\textbf{M}\in {\mathbb{R}}^{{(J+1)}\times {d}_{m}}$ represents primary feature, and $\textbf{L}\in {\mathbb{R}}^{{(J+1)}\times {d}_{m}}$ represents auxiliary feature. Meanwhile, \textbf{Q}, \textbf{K} and \textbf{V} are divided into ${n}$ attention heads.

\subsubsection{\textbf{Adaptive Adjustment Mechanism}} \label{secC.2}
To reduce reliance on visual features and align modality features, we developed an AAM, illustrated in Fig. \ref{structure}. The mechanism is implemented via a teacher-student model, where Faster R-CNN~\cite{ren2015faster}, trained on the UAV RGB dataset~\cite{munir2024investigation_rgb_dtc}, serves as the teacher network. 
Alignment between visual and audio features is achieved by correlating the drone's position across both modalities. Furthermore, the prediction acts as an adaptive adjustment factor $ \alpha$, modulating the relative contributions of visual and audio features. 
This enables adaptive feature fusion, allowing visual features to support audio-based predictions without overshadowing them. The calculation is as follows:

\vspace{-1em}
\begin{equation} 
\begin{matrix}
\begin{aligned}
\textbf{$\digamma $} 
 &=\text{$\Phi$} ({\textbf{$\Psi$}}_{T}, {\textbf{$\Psi$}}_{S}) +  \text{$\Phi$} (\text{$\Phi$}({\textbf{$\Psi$}}_{T}, {\textbf{$\Psi$}}_{S}),\alpha* {\textbf{$\Upsilon$}}),
\end{aligned}
\end{matrix}
\end{equation}
\vspace{-1.5em}

Where  ${\textbf{$\Upsilon$}}$ refers to the visual feature. $\alpha$ denotes the probability of the existence of a drone based on visual features. 
${\textbf{$\Psi$}}_{T}$ represents the temporal feature of the audio, ${\textbf{$\Psi$}}_{S}$ denotes the spectral feature of the audio. \textbf{$\digamma $} denotes the audio-visual fusion feature.


\subsection{Prediction Head}  \label{sec3.D}
The learnable temporal token ${\textbf{t}}_{T}$ is extracted from the fused feature representation \textbf{$\digamma $}. This token is subsequently fed into a prediction head responsible for UAV trajectory estimation and classification. The prediction head is composed of two main parts: a trajectory prediction head and a drone classification head. Both components utilize multi-layer perceptrons to generate their outputs.

\begin{table*}[t]
\centering
\caption{3D Trajectory Estimations Comparison.}
\vspace{-5pt}
\label{compare_methods}
\renewcommand{\arraystretch}{1.5}

\begin{tabular}{lcccccccccccccccc}

\toprule 
\hline

\multirow{2}{*}{Modality}&\multirow{2}{*}{Network} & \multicolumn{5}{c}{Light}  & \multicolumn{5}{c}{Dark} & \multirow{2}{*}{$\overline{\text{APE}}(m)$} &
 \multirow{2}{*}{$\overline{\text{Acc}}$(\%)} \\
 
\cmidrule(lr){3-7} \cmidrule(lr){8-12}
& & $D_x$ & $D_y$ & $D_z$ &$\text{APE}$& $\text{Acc(\%)}$ & $D_x$ & $D_y$ & $D_z$ &$\text{APE}$& $\text{Acc(\%)}$\\
\midrule

\multirow{2}{*}{Visual} & VisualNet~\cite{yang2023av_audio_img_fusion} & 0.24 &  0.39 & 0.32 &  0.65 & \underline{99.7}  & 1.98  & 6.10 & 8.13  & 11.45 & 11.3 & 6.05 & 55.5 \\ 

 & DarkNet~\cite{bochkovskiy2020yolov4}  & \underline{0.23}  & 0.46  & \textbf{0.23}  & \underline{0.63}  & \textbf{100}  & 1.84   & 5.50  & 4.57 & 8.31  & 25.9 & 4.47 & 63.0  \\

\hline

\multirow{3}{*}{Audio}& AudioNet~\cite{yang2023av_audio_img_fusion}  & 0.60   & 1.76   & 1.59  & 2.80  &  79.8  & 0.60  &1.76  &1.59   & 2.80&79.8 & 2.80& 79.8  \\

 & DroneChase~\cite{vora2023dronechase} & 0.54  & 1.59  & 1.51  & 2.64  & 80.6  & \underline{0.54}  & 1.59 & 1.51  & 2.64 & 80.6 & 2.64 & 80.6  \\

    \hline

\multirow{3}{*}{Audio-Visual} & TalkNet ~\cite{tao2021someone}  & 0.31  & 0.69  & 0.44  & 0.99  &  \textbf{100}  & 1.13  & 3.39 & 3.92  & 5.82 & 47.4 & 3.41 &73.7  \\

& AV-ped~\cite{yang2023av_audio_img_fusion} &  0.31 & 0.50  & 0.59  & 0.97  & 98.5  & 0.58   & 1.54  & 2.26 & 3.13  & 80.7 & 2.01&89.6  \\

& AV-FDTI~\cite{yang2024av_Audio-visual} &\textbf{ 0.13}  &  \underline{0.31} & 0.38  &\textbf{ 0.58 } & 99.6   & \textbf{0.35}  & \underline{1.06} & \underline{1.10}  & \underline{1.89}& \underline{88.3} & \underline{1.24} & \underline{94.0}  \\





&\textbf{AV-DTEC}(Ours)  & 0.33  & \textbf{0.25}  & \underline{0.27}  &\textbf{0.58}   & \underline{99.7}  &\textbf{ 0.35}  & \textbf{0.35}  & \textbf{0.38}  &\textbf{0.75}   & \textbf{98.9}  &\textbf{0.67}   & \textbf{99.3}   \\

\hline 
    \bottomrule
\end{tabular}
\vspace{2pt}\\
\footnotesize{Best results are in $\textbf{bold}$, and second best in $\underline{underline}$. $\overline{Overline}$: the mean value of day and night.}
\vspace{-2.em}
\end{table*}

\textbf{Trajectory Head:} The audio often contains various noises, which can lead to errors in trajectory prediction. To enhance the model's robustness against noise during training, we employ the L1 loss function. This choice helps stabilize the model by reducing the impact of noise and preventing significant deviations from the expected trajectory. The loss function is defined as follows:
\vspace{-0.5em}
\begin{equation} 
\begin{matrix}
\begin{aligned}
{L}_{pos}=\frac{1}{N}\displaystyle\sum_{i=1}^{N}\left | \hat{{O}_{i}}-{o}_{i}\right |,
\end{aligned}
\end{matrix}
\end{equation}
\vspace{-1em}

where \textit{N} denotes the total number of UAVs, $\hat{O}$ represents the label generated by LiDAR, and ${o}$ denotes the predicted trajectory.

\textbf{Classification Head:} 
Attributes of the UAV, such as its size and audio signature, provide key insights into its category, which is vital for anti-UAV detection systems. Accurate classification enhances the system's ability to locate the drone's 3D coordinates more precisely and implement appropriate countermeasures. To address this need, a classification head is designed to classify the UAV. This head utilizes cross-entropy loss to optimize the classification, and the loss function is defined as follows:
\vspace{-0.5em}
\begin{equation} 
\begin{matrix}
\begin{aligned}
{L}_{cls}=-\frac{1}{N}\displaystyle\sum_{i=1}^{N}{\Lambda}_{i}log({\lambda}_{i}),
\end{aligned}
\end{matrix}
\end{equation}
where \textit{N} denotes the total number of UAV category, ${\Lambda}_{i}$ represents the ground truth class, and ${\lambda}_{i}$ represents the predicted class.

The loss function for the joint teacher-student model mirrors the loss used in the prediction head. However, the key distinction lies in using a binary cross-entropy loss in the student model, which signifies the presence or absence of a UAV in the visual. Additionally, when a UAV is detected, the position loss is preserved. This combination ensures that the model not only recognizes the drone but also retains precise positioning information.

Therefore, the overall training loss function is given by:

\vspace{-1em}
\begin{equation} 
\begin{matrix}

\begin{aligned}

{L}_{total} = {L}_{cls} + {\gamma }_{1}{L}_{pos} + {\gamma }_{2}{L}_{t-s},
\end{aligned}

\end{matrix}
\end{equation}
\vspace{-1.5em}

where $ {\gamma }_{1}$ and $ {\gamma }_{2}$ are the balancing factor for the multi-task loss. $ {L}_{t-s}$ denotes teacher-student model loss which also consist of $ {L}_{cls}$  and ${L}_{pos}$.

\subsection{Unsupervised Pseudo Label Generation}  \label{sec3.E}
In this work, we propose a self-supervised AV-DETC using pseudo-labels generated by an unsupervised LiDAR-based trajectory estimation framework. The framework employs DBSCAN clustering to segment LiDAR data, a method widely used in anti-UAV applications \cite{deng2024multi_point, xiao2024clustering} due to its accuracy. We extend the existing approach by extracting both spatial and temporal clusters and empirically filtering outliers based on point size, shape, and spatial-temporal continuity. While LiDAR offers precise spatial information, its high cost and data intensity limit scalability. To address this, we use LiDAR to generate pseudo-labels with up to 1-meter accuracy, which is then used to train a more cost-effective visual-audio network. The parameters for the network are chosen empirically. This hybrid approach combines the accuracy of LiDAR with the scalability and affordability of visual and audio sensors, offering a balanced solution for UAV detection.


\begin{figure}[t]
\centering
\includegraphics[width=8cm]{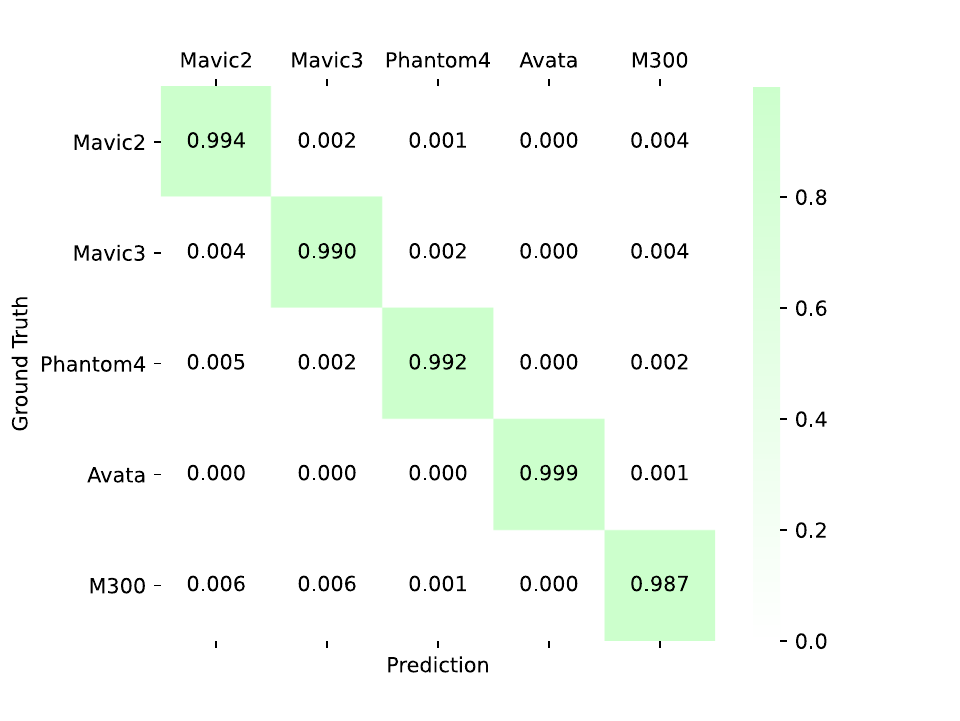}
\vspace{-1em}
\caption{The $\overline{\text{Acc}}$ confusion matrix for the classification results of AV-DTEC.}
\label{av-conf} 
 \vspace{-1em}
\end{figure}

\begin{figure}[t]
\centering
\includegraphics[width=8.6cm]{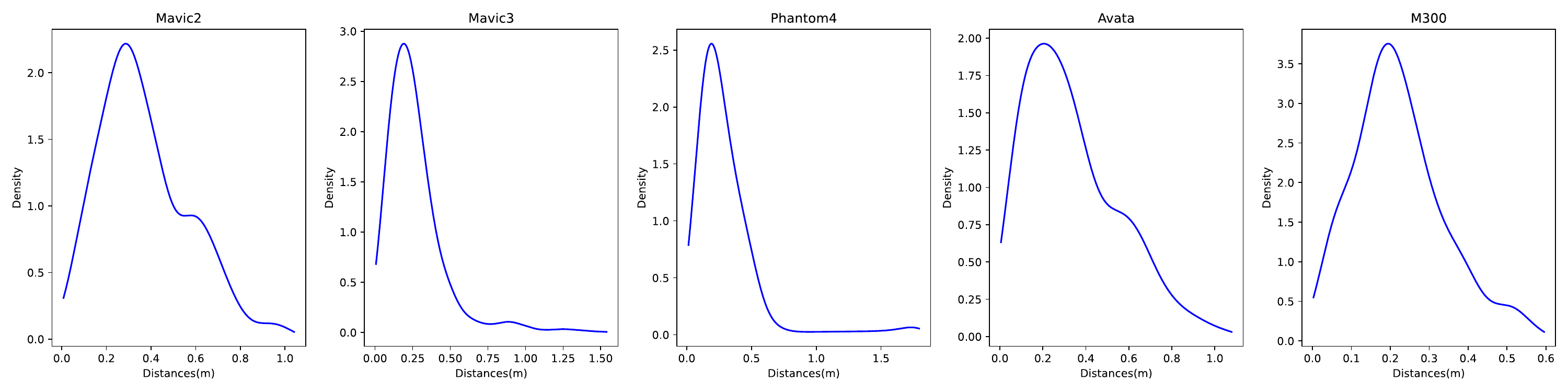}
\vspace{-1.5em}
\caption{The difference distribution between pseudo labels generated by LiDAR and ground truth.}
\label{distribution} 
 \vspace{-2em}
\end{figure}

\section{EXPERIMENT}
\subsection{Dateset}
We use the MMAUD anti-UAV dataset~\cite{yuan2024mmaud}. The dataset includes real-world images, LiDAR, millimeter-wave point clouds, four-channel audio, and survey-graded ground truth. 
The dataset is divided into training and testing datasets with a ratio of 7:3.
\subsection{Experimental Setting}
\textbf{Implementation Detail:} The model is trained using the Adam optimizer with a batch size of 64, an initial learning rate of 0.0001, and 200 epochs. All experiments are conducted on an NVIDIA GeForce RTX 4090 GPU. For the audio patch split, we set \textit{w}=4 and \textit{h}=1.  For image patch split, we set \textit{w}=16 and \textit{h}=16. For the AMamba, we set\textit{ L} = 12. For the FEM, we set ${n}$ = 6, ${d}_{m}$ = ${d}_{k}$ = 192. The multi-task balance factor is set to ${{\gamma}_{1}}$=2 and ${{\gamma}_{2}}$=0.5. To compare with other multi-modal model experiments, we apply brightness attenuation augmentation during training and testing.

\begin{figure*}[t]
\centering
\centering
\includegraphics[width=0.9\linewidth]{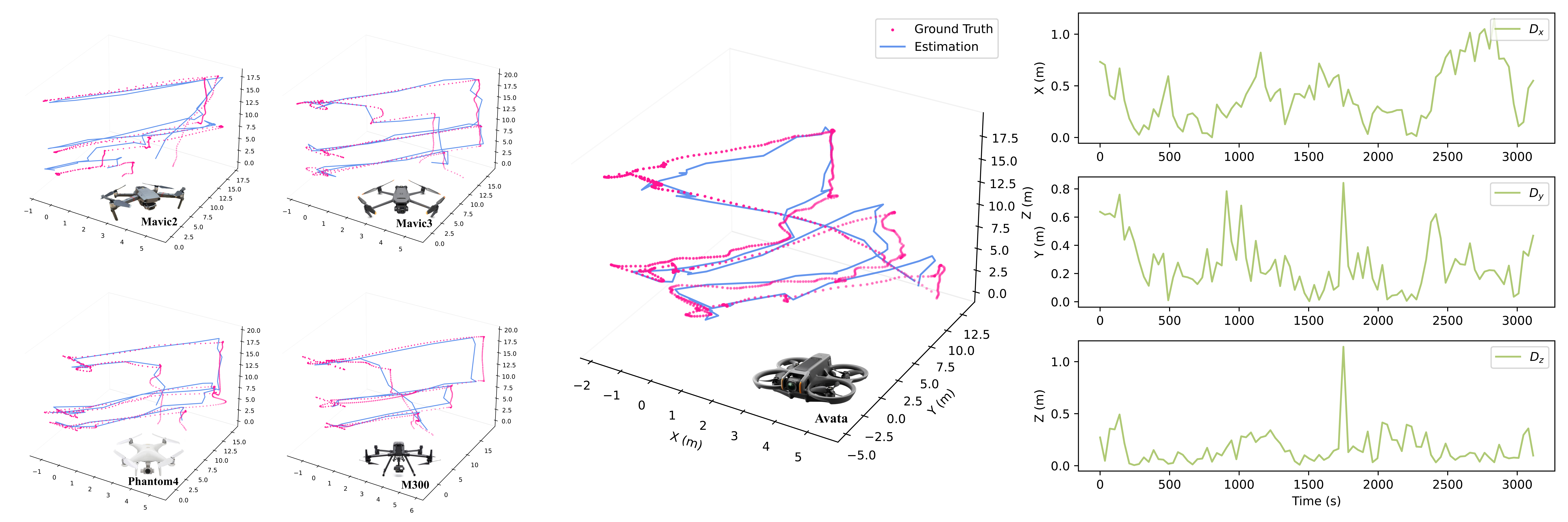}
\caption{Test set trajectory estimation: Red curves represent ground truth, blue curves show predicted trajectories.}
\label{visual_gt_pred} 
\end{figure*}

\textbf{Evaluation Metrics: }For trajectory estimation, we apply the L1 norm to measure center distances (${D}{x}$, ${D}{y}$, ${D}_{z}$) and use the Average Trajectory Error (APE) metric. For classification performance, we evaluate using accuracy (Acc).





\subsection{Baseline Selections}
To validate the effectiveness of our model, we compare it against various anti-UAV methods, including those visual~\cite{yang2023av_audio_img_fusion, bochkovskiy2020yolov4}, audio~\cite{tao2021someone, yang2023av_audio_img_fusion}, and audio-visual fusion~\cite{vora2023dronechase, yang2023av_audio_img_fusion, yang2024av_Audio-visual}.








\begin{table}
\centering
\caption{Comparison of GFLOPs and Params based on Audio-Visual fusion model.}
\label{gflops}
\renewcommand{\arraystretch}{1.5}
\begin{tabular}{lccc}
\hline
Network & GFLOPs & Params(M) \\
\hline
TalkNet &\textbf{1.53}  &  \underline{15.80}   \\
AV-ped & 5.67  &  32.36   \\
AV-FDTI  & 11.37  &  92.28   \\
\textbf{AV-DTEC}(Ours)& \underline{1.98}  &\textbf{10.16}     \\

\hline
\end{tabular}
\vspace{-1em}
\end{table}

\begin{table}[h]
\centering
\caption{The ablation study of AMamba with different modules and feature fusion.}
\label{tsmamba}
\renewcommand{\arraystretch}{1.5}
\begin{tabular}{lcccccc}
\hline
TMamba & SMamba & Concate & FEM  & $\overline{\text{APE}}(m)$& $\overline{\text{Acc}}(\%)$\\
\hline
$\checkmark$ &   &    &    & {0.89} &  97.4  & \\
&$\checkmark$   &    &    & 1.16 & 95.2 &   \\
$\checkmark$& $\checkmark$  & $\checkmark$    &  & 1.12 & 96.8 &  \\
$\checkmark$& $\checkmark$     &  & $\checkmark$ & \textbf{0.86} & \textbf{97.9} &   \\

\hline
\end{tabular}
\end{table}

\begin{table*}[t]
\centering
\caption{The ablation study of AV-DTEC with different
modules and feature fusion.}
\label{av-detc}
\renewcommand{\arraystretch}{1.5}

\begin{tabular}{lccccccccccc}

\toprule 
\hline

\multirow{2}{*}{AMamba}&
\multirow{2}{*}{Vim} & 

\multirow{2}{*}{Concate} & 
\multirow{2}{*}{FEM} & 
\multirow{2}{*}{AAM} & 

\multicolumn{2}{c}{Light}  & 
\multicolumn{2}{c}{Dark} & 
\multirow{2}{*}{$\overline{\text{APE}}(m)$} &
\multirow{2}{*}{$\overline{\text{Acc}}$(\%)}\\

 
\cmidrule(lr){6-7} \cmidrule(lr){8-9}
& & & & & $\text{APE}$& $\text{Acc(\%)}$ & $\text{APE}$& $\text{Acc(\%)}$\\
\midrule

 $\checkmark$ &  & & &   &{0.86}   & 97.9  &\underline{0.86}   & \text{97.9}  & {0.86}   & \text{97.9}   \\

 $\checkmark$ & $\checkmark$  &  $\checkmark$ && &  \textbf{0.52}   & \underline{99.6}  &\textbf{0.75}   & \underline{98.2}  &\textbf{0.64}   & \underline{98.9}   \\

$\checkmark$ & $\checkmark$  && $\checkmark$&& 0.60   & 99  &\text{0.89}   & \text{97.8}  &\text{0.75}   & {98.4} &  \\

$\checkmark$ & $\checkmark$ && $\checkmark$& $\checkmark$& \underline{0.58 }  &\textbf{99.7}  &\textbf{0.75}   & \textbf{98.9}  &\underline{0.67}   & \textbf{99.3}
\\

\hline 
    \bottomrule
\end{tabular}
\end{table*}

Table \ref{compare_methods} presents the experimental results for each network. All other networks are supervised learning except for our AV-DTEC. AV-DTEC has achieved state-of-the-art results in $\overline{\text{APE}}$ and $\overline{\text{Acc}}$. For single-modality trajectory estimation and classification, although the accuracy of visual-based networks is generally better than audio-based networks during the day, their performance rapidly degrades when the illumination changes. This is unacceptable in anti-UAV systems. However, the audio-based networks do not change. This shows that audio is more suitable for UAV detection. The audio-visual fusion model just combines the advantages of the two modalities. The accuracy during the day is improved, while the impact of illumination changes is alleviated. However, there is still a large degree of attenuation. The main reason is that the visual fusion framework does not fully extract the features of audio and over-relies the contribution of visual features. Other audio-visual fusion networks feature extraction backbones, which all use convolution neural network (CNN), which fails to capture the global temporal characteristics of audio due to the local attention mechanism. Thus they cannot learn temporal difference information. AV-DTEC's selective scanning aligns well with the sequence features in the time dimension of the audio spectrogram. Therefore, AV-DTEC is better suited for extracting audio spectrogram features. To aid in understanding, we provide a visualization of UAV trajectory estimation in Fig. \ref{visual_gt_pred} and classification in Fig. \ref{av-conf}.

Table \ref{gflops} shows the resource consumption experiment of the audio-visual network. It can be seen that AV-DTEC achieves the best performance while minimizing the number of parameters. Although TalkNet has the smallest floating point operations, its overall performance lags behind AV-DTEC by more than 20\%. Therefore, AV-DTEC is more suitable for deployment on mobile devices.

\subsection{Ablation Study and Analysis}  \label{sec4.ablation}

In this section, we perform ablation experiments to evaluate the AV-DTEC effectiveness in feature extraction including AMamba, FEM, and AAM. All experiments are conducted in a self-supervised.
\subsubsection{AMamba}   
 AMamba consists of TMamba and SMamba. As shown in Table \ref{tsmamba}, both TMamba and SMamba surpass the audio-based network in Table \ref{compare_methods}. This shows that the selective SSM is more suitable for extracting the temporal feature of audio. In addition, the performance of TMamba is better than that of SMamba, indicating that TMamba extracts more sound propagation features. Therefore, we use the temporal feature extracted by TMamba as the primary feature and the spectral feature extracted by SMamba as the auxiliary feature and enhance them using FEM other than concatenate directly. The highest performance is obtained based on audio, which also effectively proves the effectiveness of feature enhancement by FEM.

\subsubsection{AV-DECT}   
The comparative experiments in Table \ref{compare_methods} show that the detection based on audio feature is more stable than that based on visual feature, and the performance of the modalities fusion model is generally higher than that of a single modality. Therefore, in order to improve the performance of the model, we use FEM to integrate visual features into audio features. As can be seen in Table \ref{av-detc}. Using FEM directly has lower performance than concatenate. The main reason is that there is an error between the pseudo label and the ground truth as shown in Fig. \ref{distribution}. FEM makes the correlation between modalities stronger, and there is also a problem of misalignment between modalities. When the visual feature fails, it has a greater impact on the results. When AAM is introduced, the teacher-student model solves the problem of alignment between modalities during training, and at the same time reversely weakens the weight of the visual feature based on its own predicted results during inference. Therefore, when the illumination changes, the performance of AV-DTEC is less degraded.

\section{Conclusion}
In this work, we propose an innovative self-supervised audio-visual model, AV-DECT, for detecting UAV threats. AV-DECT treats the audio feature as the primary feature and integrates visual features through the designed feature enhancement module. Additionally, AV-DECT incorporates an adaptive adjustment mechanism to align modalities and weakens the reliance on the visual feature. This approach achieves optimal performance in both UAV trajectory estimation and classification, offering a robust solution for anti-UAV systems.


\bibliographystyle{IEEEtran}
\bibliography{mybib}


\end{document}